\PassOptionsToPackage{unicode}{hyperref}
\PassOptionsToPackage{hyphens}{url}
\PassOptionsToPackage{dvipsnames,svgnames,x11names}{xcolor}
\documentclass[
]{article}
\usepackage{amsmath,amssymb}
\usepackage{lmodern}
\usepackage{iftex}
\ifPDFTeX
  \usepackage[T1]{fontenc}
  \usepackage[utf8]{inputenc}
  \usepackage{textcomp} % provide euro and other symbols
\else % if luatex or xetex
  \usepackage{unicode-math}
  \defaultfontfeatures{Scale=MatchLowercase}
  \defaultfontfeatures[\rmfamily]{Ligatures=TeX,Scale=1}
\fi
\IfFileExists{upquote.sty}{\usepackage{upquote}}{}
\IfFileExists{microtype.sty}{% use microtype if available
  \usepackage[]{microtype}
  \UseMicrotypeSet[protrusion]{basicmath} % disable protrusion for tt fonts
}{}
\makeatletter
\@ifundefined{KOMAClassName}{% if non-KOMA class
  \IfFileExists{parskip.sty}{%
    \usepackage{parskip}
  }{% else
    \setlength{\parindent}{0pt}
    \setlength{\parskip}{6pt plus 2pt minus 1pt}}
}{% if KOMA class
  \KOMAoptions{parskip=half}}
\makeatother
\usepackage{xcolor}
\usepackage{graphicx}
\makeatletter
\def\maxwidth{\ifdim\Gin@nat@width>\linewidth\linewidth\else\Gin@nat@width\fi}
\def\maxheight{\ifdim\Gin@nat@height>\textheight\textheight\else\Gin@nat@height\fi}
\makeatother
\setkeys{Gin}{width=\maxwidth,height=\maxheight,keepaspectratio}
\makeatletter
\def\fps@figure{htbp}
\makeatother
\NewDocumentCommand\citeproctext{}{}
\NewDocumentCommand\citeproc{mm}{%
  \begingroup\def\citeproctext{#2}\cite{#1}\endgroup}
\makeatletter
 \let\@cite@ofmt\@firstofone
 \def\@biblabel#1{}
 \def\@cite#1#2{{#1\if@tempswa , #2\fi}}
\makeatother
\newlength{\cslhangindent}
\newlength{\csllabelwidth}
\newenvironment{CSLReferences}[2] % #1 hanging-indent, #2 entry-spacing
 {\begin{list}{}{%
  \setlength{\itemindent}{0pt}
  \setlength{\leftmargin}{0pt}
  \setlength{\parsep}{0pt}
  \ifodd #1
   \setlength{\leftmargin}{\cslhangindent}
   \setlength{\itemindent}{-1\cslhangindent}
  \fi
  \setlength{\itemsep}{#2\baselineskip}}}
 {\end{list}}
\usepackage{calc}

\ifLuaTeX
\usepackage[bidi=basic]{babel}
\else
\usepackage[bidi=default]{babel}
\fi
\babelprovide[main,import]{american}
\def\languageshorthands#1{}
\ifLuaTeX
  \usepackage{selnolig}  % disable illegal ligatures
\fi
\IfFileExists{bookmark.sty}{\usepackage{bookmark}}{\usepackage{hyperref}}
\IfFileExists{xurl.sty}{\usepackage{xurl}}{} % add URL line breaks if available
\hypersetup{
  pdftitle={cosmo-numba: B-modes and COSEBIs computations accelerated by
Numba},
  pdfauthor={Axel Guinot, Rachel Mandelbaum},
  pdflang={en-US},
  colorlinks=true,
  linkcolor={Maroon},
  filecolor={Maroon},
  citecolor={Blue},
  urlcolor={Blue},
  pdfcreator={LaTeX via pandoc}}

\title{cosmo-numba: B-modes and COSEBIs computations accelerated by
Numba}

\definecolor{c53baa1}{RGB}{83,186,161}
\definecolor{c202826}{RGB}{32,40,38}

\usepackage[affil-it]{authblk}
\usepackage{orcidlink}
\author[1%
  ]{Axel Guinot%
    \,\orcidlink{0000-0002-5068-7918}\,%
    }
\author[1%
  ]{Rachel Mandelbaum%
    \,\orcidlink{0000-0003-2271-1527}\,%
    }

\affil[1]{Department of Physics, McWilliams Center for Cosmology and
Astrophysics, Carnegie Mellon University, Pittsburgh, PA 15213, USA%
  }
\date{6 March 2026}

\begin{document}
\maketitle

\section{Summary}\label{summary}

Weak gravitational lensing is a widely used probe in cosmological
analysis. It allows astrophysists to understand the content and
evolution of the Universe. We are entering an era where we are not
limited by the data volume but by systematic uncertainties. It is in
this context that we present here a simple python-based software package
to help in the computation of E-/B-mode decomposition, which can be use
for systematic checks or science analysis. As we demonstrate, our
implementation has both the high precision and speed required to perform
this kind of analysis while avoiding a scenario wherein either numerical
precision or computational time is a significant limiting factor.

\section{Statement of need}\label{statement-of-need}

The E-/B-mode composition for cosmic shear poses a significant
computational challenge given the need for high precision (required to
integrate oscillatory functions over a large integration range and
achieve accurate results) and speed. \texttt{Cosmo-numba} meets this
need, facilitating the computation of E-/B-mode decomposition using two
methods. One of them is the Complete Orthogonal Sets of E-/B-mode
Integrals (COSEBIs) as presented in P. Schneider et al.
(\citeproc{ref-Schneider_2010}{2010}). The COSEBIs rely on very high
precision computation requiring more than 80 decimal places. P.
Schneider et al. (\citeproc{ref-Schneider_2010}{2010}) propose an
implementation using \texttt{mathematica}. \texttt{cosmo-numba} uses a
combination of \texttt{sympy} and \texttt{mpmath} to reach the required
precision. This python version enables an easier integration within
cosmological inference pipelines, which are commonly python-based, and
facilitates the null tests.

This software package also enables the computation of the pure-mode
correlation functions presented in Peter Schneider et al.
(\citeproc{ref-Schneider_2022}{2022}). Those integrals are less
numerically challenging than the COSEBIs, but having a fast computation
is necessary for their integration in an inference pipeline. Indeed, one
can use those correlation functions for cosmological inference, in which
case the large number of calls to the likelihood function will require a
fast implementation.

\section{State of the field}\label{state-of-the-field}

There are other implementations of the COSEBIs such as
CosmoPipe\footnote{\url{https://github.com/AngusWright/CosmoPipe}} used
in the KiDS-legacy analysis (\citeproc{ref-kids_legacy}{Wright et al.,
2025}). Our implemetation is characterized by the use of \texttt{numba}
that makes the computation of the filter functions described in
\autoref{sec:cosebis} faster. Regarding the pure E-/B-mode
decomposition, we have not found a similar publicly available
implementation. That being said, they are classicaly used as a one-time
measure for null tests in various surveys. The implementation we are
presenting would enable one to use this decomposition for cosmological
inference, which requires computing several integrals at each likelyhood
call. While the commonly used \texttt{scipy} library would make the
computation untractable, the speed gain by switching to \texttt{numba}
opens new opportunities such as this one.

\section{Software design}\label{software-design}

This package has been designed around two constraints: precision and
speed. As it can be difficult to reach both at the same time, the code
is partitioned in a way that parts requiring high precision are done
using python libraries such as \texttt{sympy} and \texttt{mpmath}. In
contrast, parts of the code that do not require high precision leverage
the power of Just-In-Time (JIT) compilation. \texttt{Numba} provides
significant speed up compared to a classic python implementation. As
this library is intended to provide tools for cosmological computation,
it was important to provide meaningful unit tests and demonstrate a full
coverage of the library. Providing an accurate coverage is challenging
when using \texttt{numba} compiled code. Our implementation allows the
developer to disable compilation for the targeted part of the code when
performing coverage tests. This allows us to provide both high quality
unit tests and good coverage to the users.

\section{Testing setup}\label{testing-setup}

In the following two sections we make use of fiducial shear-shear
correlation functions, \(\xi_{\pm}(\theta)\), and power spectra,
\(P_{E/B}(\ell)\). They have been computed using the Core Cosmology
Library\footnote{\url{https://github.com/LSSTDESC/CCL}}
(\citeproc{ref-Chisari_2019}{Chisari et al., 2019}). The cosmological
parameters are taken from Aghanim et al.
(\citeproc{ref-Planck_2018}{2020}). For tests that involved covariances
we are using the Stage-IV Legacy Survey of Space and Time (LSST) Year 10
as a reference. The characteristics are taken from the LSST Dark Energy
Science Collaboration (DESC) Science Requirements Document (SRD)
(\citeproc{ref-LSST_SRD}{The LSST Dark Energy Science Collaboration et
al., 2021}).

\section{COSEBIs}\label{cosebis}

\label{sec:cosebis}

The COSEBIs are defined as:

\begin{equation}
E_{n} = \frac{1}{2} \int_{0}^{\infty} d\theta\, \theta [T_{n,+}(\theta)\xi_{+}(\theta) + T_{n,-}(\theta)\xi_{-}(\theta)],
\end{equation} \begin{equation}
B_{n} = \frac{1}{2} \int_{0}^{\infty} d\theta\, \theta [T_{n,+}(\theta)\xi_{+}(\theta) - T_{n,-}(\theta)\xi_{-}(\theta)];
\end{equation}

where \(\xi_{\pm}(\theta)\) are the shear correlation functions, and
\(T_{n,\pm}\) are the weight functions for the COSEBI mode \(n\). The
complexity is in the computation of the weight functions.
\texttt{Cosmo-numba} carries out the computation of the weight functions
in a logarithmic scale defined by:

\begin{equation}
T_{n,+}^{\rm{log}}(\theta) = t_{n,+}^{\rm{log}}(z) = N_{n}\sum_{j=0}^{n+1}\bar{c}_{nj}z^{j};
\end{equation}

whare \(z = \rm{log}(\theta/\theta_{\rm{min}})\), \(N_{n}\) is the
normalization for the mode \(n\), and \(\bar{c}_{jn}\) are defined
iteratively from Bessel functions (we refer the readers to P. Schneider
et al. (\citeproc{ref-Schneider_2010}{2010}) for more details).

\begin{figure}
\centering
\includegraphics[keepaspectratio]{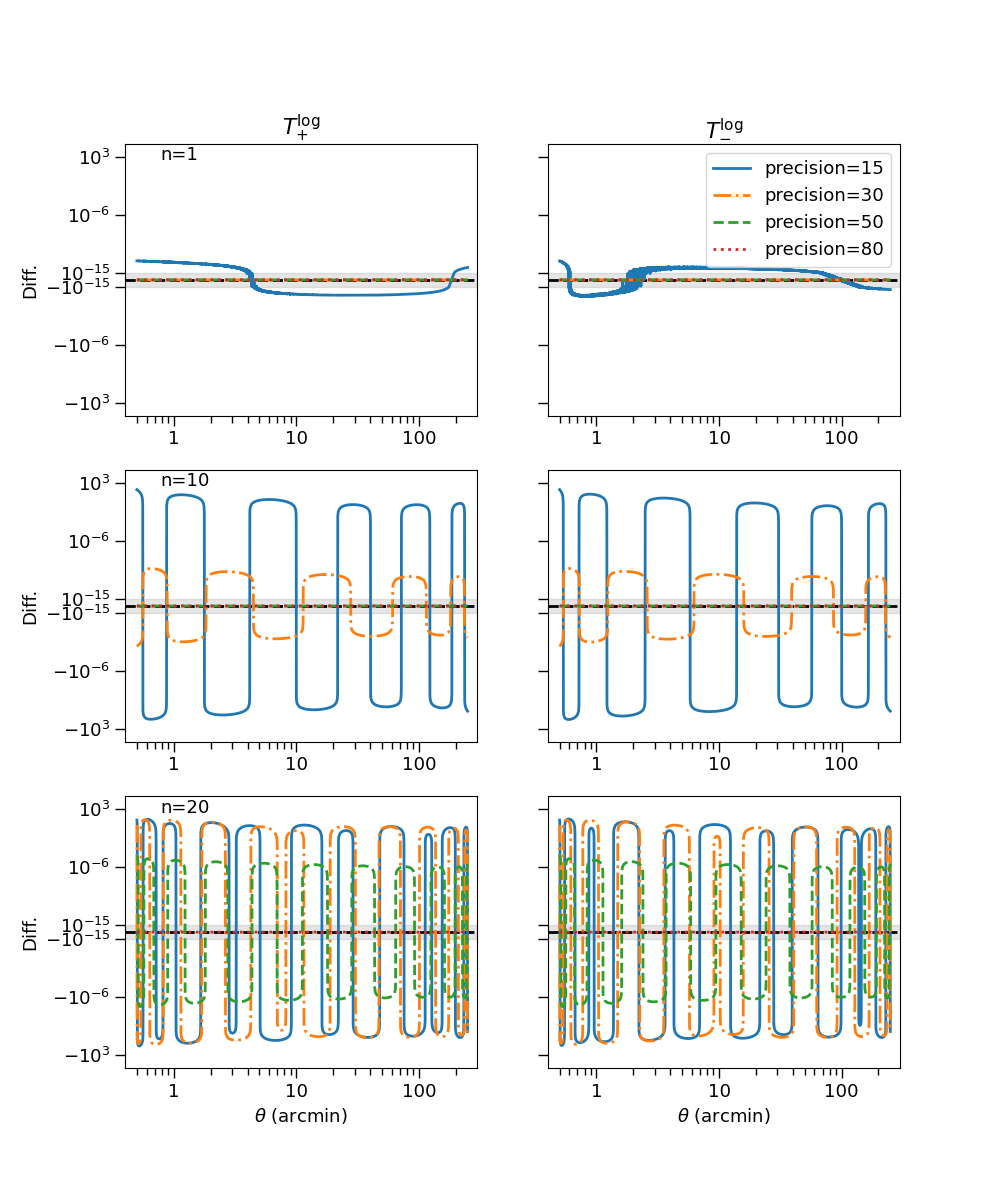}
\caption{In this figure we show the impact of the precision in the
computation of the weight functions \(T_{\pm}^{\rm{log}}\). For
comparion, a precision of 15 corresponds to what would be achieved using
\texttt{numpy} float64. The difference is computed with respect to the
original \texttt{Mathematica} implementation presented in P. Schneider
et al. (\citeproc{ref-Schneider_2010}{2010}). The figure uses symlog,
with the shaded region representing the linear scale in the range
\([-10^{-15}, 10^{-15}]\).\label{fig:Tpm_prec}}
\end{figure}

\begin{figure}
\centering
\includegraphics[keepaspectratio]{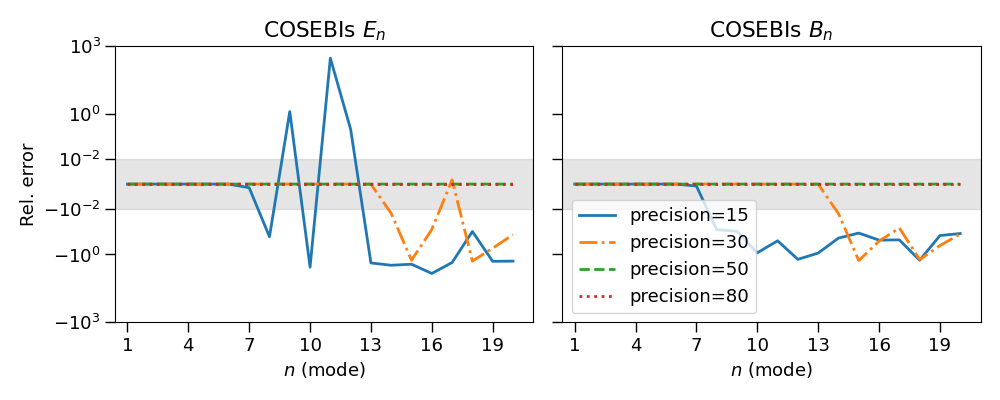}
\caption{This figure shows the difference in the COSEBIs E- and B-modes
relative to the original \texttt{Mathematica} implementation. We see
that using only 15 decimal places would lead to several percent error,
making an implementation based on \texttt{numpy} float64 not suitable.
The figure uses symlog, with the shaded region representing the linear
scale in the range \([-1,1]\) percent.\label{fig:EB_prec}}
\end{figure}

We have validating our implementation against the original version in
\texttt{Mathematica} from P. Schneider et al.
(\citeproc{ref-Schneider_2010}{2010}). In \autoref{fig:Tpm_prec} we show
the impact of the precision going from 15 decimal places, which
corresponds to the precision one could achieve using float64, up to 80
decimal places, the precision used in the original \texttt{Mathematica}
implementation. We can see that classic float64 precision would not be
sufficient, and with a precision of 80 our code exactly recovers the
results from the original implementation. Given that the precision comes
at very little computational cost, we default to the original
implementation using high precision. The impact of the precision
propagated to the COSEBIs is shown in \autoref{fig:EB_prec}. We can see
that using a lower precision than the default setting can incur a
several percent error.

\begin{figure}
\centering
\includegraphics[keepaspectratio]{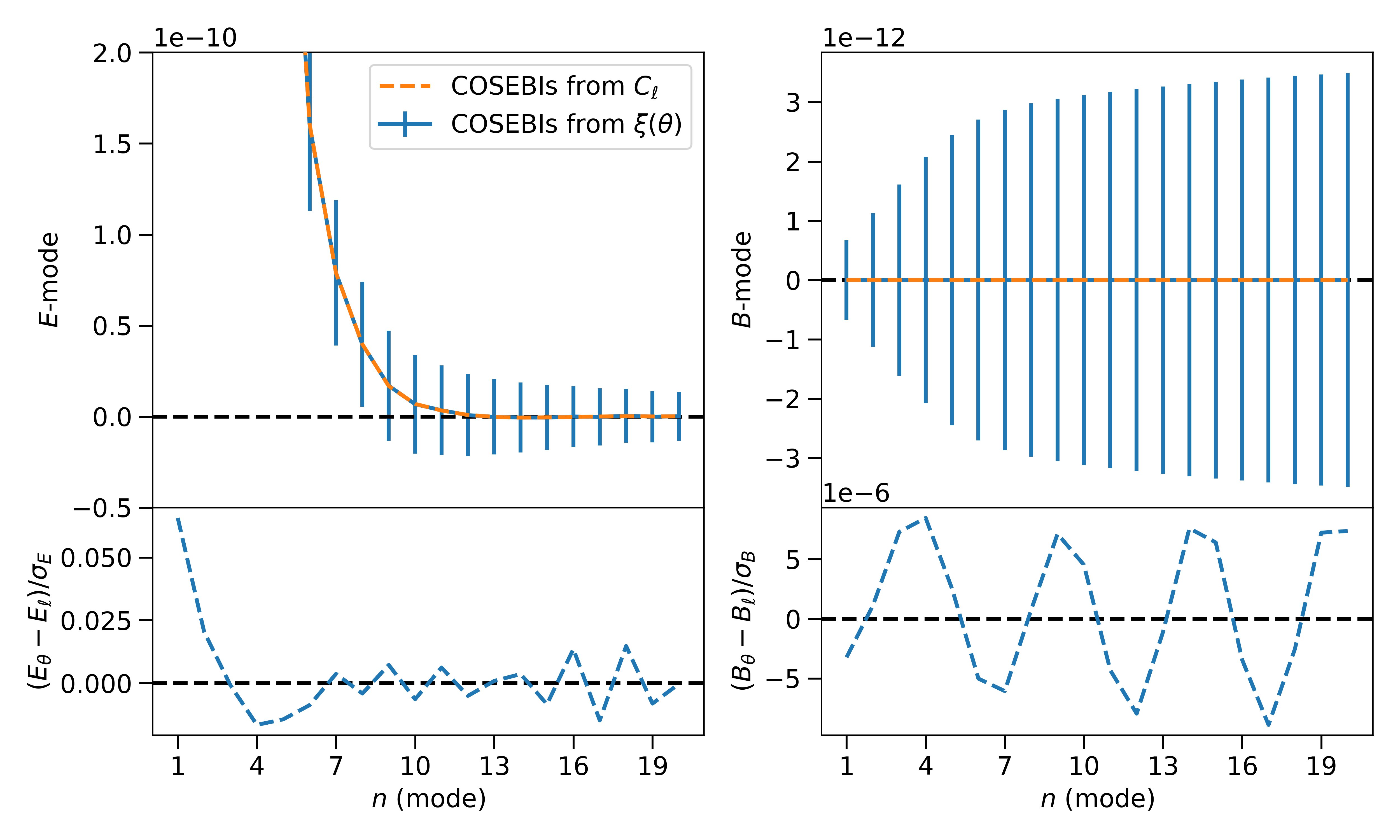}
\caption{Comparison of the COSEBIs E- and B-mode computed from
\(\xi_{\pm}(\theta)\) and \(C_{E/B}(\ell)\). The \emph{upper} panel
shows the COSEBIs E-/B-modes while the \emph{bottom} panel shows the
difference with respect to the statistical uncertainty based on the LSST
Y10 covariance.\label{fig:cosebis_xi_cl}}
\end{figure}

COSEBIs can also be defined from the power spectrum as:

\begin{equation}
E_{n} = \int_{0}^{\infty} \frac{d \ell \, \ell}{2 \pi} P_{E}(\ell)W_{n}(\ell);
\end{equation} \begin{equation}
B_{n} = \int_{0}^{\infty} \frac{d \ell \, \ell}{2 \pi} P_{B}(\ell)W_{n}(\ell);
\end{equation}

where \(P_{E/B}(\ell)\) is the power spectrum of E- and B-modes and
\(W_{n}(\ell)\) are the filter functions which can be computed from
\(T_{n,+}\) as:

\begin{equation}\label{eq:Wn}
W_{n}(\ell) = \int_{\theta_{\rm{min}}}^{\theta_{\rm{max}}} d\theta \,\theta T_{n,+}(\theta) J_{0}(\ell\theta);
\end{equation}

with \(J_{0}(\ell \theta)\) the 0-th order Bessel function. The
\autoref{eq:Wn} is a Hankel transform of order 0. It can be computed
using the \texttt{FFTLog} algorithm presented in Hamilton
(\citeproc{ref-Hamilton_2000}{2000}) implemented here in \texttt{Numba}.
\autoref{fig:cosebis_xi_cl} shows the comparison between the COSEBIs
computed from \(\xi_{\pm}(\theta)\) and from \(C_{E/B}(\ell)\). We can
see that the COSEBI E- \& B-modes agree very well, with at most
\(0.3\sigma\) difference with respect to the LSST Y10 covariance. We
consider that using either approach would not impact the scientific
interpretation and both could be used for consistency checks.

Finally, we have compared our implementation agains CosmoPipe\footnote{The
  test has been done on a Mac M3 Max using 16 cores. The script to run
  the test is available at:
  \url{https://github.com/aguinot/cosmo-numba/blob/main/notebooks/cosebis_comparison_with_cosmopipe.ipynb}}
which make use of a different integration method to compute the filter
functions such as Levin integration. We found that our implementation
using numba is around 100 times faster for equivalent precision.

\section{Pure-Mode Correlation
Functions}\label{pure-mode-correlation-functions}

\begin{figure}
\centering
\includegraphics[keepaspectratio]{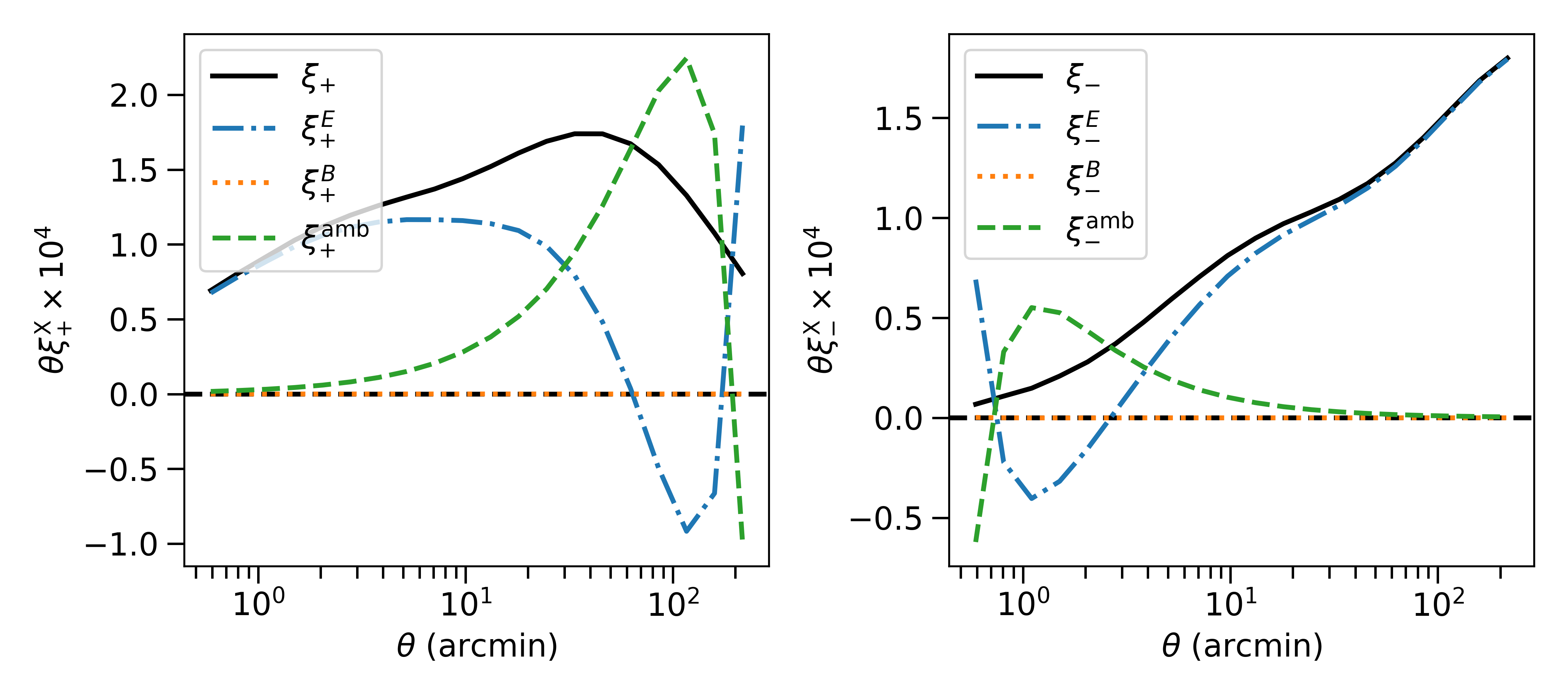}
\caption{This figure shows the decomposition of the shear-shear
correlation functions into E- and B-modes (and ambiguous
mode).\label{fig:pure_EB}}
\end{figure}

In this section we describe the computation of the pure-mode correlation
functions as defined in Peter Schneider et al.
(\citeproc{ref-Schneider_2022}{2022}). These are defined as follows:

\begin{equation}
\xi_{+}^{E}(\vartheta) = \frac{1}{2} \left[ \xi_{+}(\vartheta) + \xi_{-}(\vartheta) + \int_{\vartheta}^{\vartheta_{\rm{max}}} \frac{d \theta}{\theta} \xi_{-}(\theta) \left( 4 - \frac{12\vartheta^{2}}{\theta^{2}} \right) \right] - \frac{1}{2} \left[ S_{+}(\vartheta) + S_{-}(\vartheta)\right],
\end{equation} \begin{equation}
\xi_{+}^{B}(\vartheta) = \frac{1}{2} \left[ \xi_{+}(\vartheta) - \xi_{-}(\vartheta) - \int_{\vartheta}^{\vartheta_{\rm{max}}} \frac{d \theta}{\theta} \xi_{-}(\theta) \left( 4 - \frac{12\vartheta^{2}}{\theta^{2}} \right) \right] - \frac{1}{2} \left[ S_{+}(\vartheta) - S_{-}(\vartheta)\right],
\end{equation}

\begin{equation}
\xi_{-}^{E}(\vartheta) = \frac{1}{2} \left[ \xi_{+}(\vartheta) + \xi_{-}(\vartheta) + \int_{\vartheta_{\rm{min}}}^{\vartheta} \frac{d \theta\~\theta}{\vartheta^{2}} \xi_{+}(\theta) \left( 4 - \frac{12\theta^{2}}{\vartheta^{2}} \right) \right] - \frac{1}{2} \left[ V_{+}(\vartheta) + V_{-}(\vartheta)\right],
\end{equation} \begin{equation}
\xi_{-}^{B}(\vartheta) = \frac{1}{2} \left[ \xi_{+}(\vartheta) - \xi_{-}(\vartheta) + \int_{\vartheta_{\rm{min}}}^{\vartheta} \frac{d \theta\~\theta}{\vartheta^{2}} \xi_{+}(\theta) \left( 4 - \frac{12\theta^{2}}{\vartheta^{2}} \right) \right] - \frac{1}{2} \left[ V_{+}(\vartheta) - V_{-}(\vartheta)\right];
\end{equation}

where \(\xi_{\pm}(\theta)\) correspond to the shear-shear correlation
function. The functions \(S_{\pm}(\theta)\) and \(V_{\pm}(\theta)\) are
themselves defined by integrals and we refer the reader to Peter
Schneider et al. (\citeproc{ref-Schneider_2022}{2022}) for more details
about their definition. By contrast with the computation of the COSEBIs,
these integrals are more stable and straightforward to compute but still
require some level of precision. This is why we are using the
\texttt{qags} method from QUADPACK
(\citeproc{ref-piessens2012quadpack}{Piessens et al., 2012}) with a 5th
order spline interpolation. In addition, as one can see from the
equations above, the implementation requires a loop over a range of
\(\vartheta\) values. This is why having a fast implementation will be
required if one wants to use those correlation functions in cosmological
inference. In \autoref{fig:pure_EB} we show the decomposition of the
shear-shear correlation function into the E-/B-modes correlation
functions and ambiguous mode.

To assess the speed improvement of our implementation, we have run the
same computation using \texttt{Scipy} functions: \texttt{CubicSpline}
for the interpolation and \texttt{quad} for the integration\footnote{The
  test has been done on a Mac M3 Max using 16 cores. The script to run
  the test is available at:
  \url{https://github.com/aguinot/cosmo-numba/blob/main/notebooks/pure_EB_modes_comparison_with_scipy.ipynb}}.
While the precision is comparable, our serial version is more than 8
time faster while the parallel version is more than 50 times faster.

\section{Research impact statement}\label{research-impact-statement}

This software is being in the Utraviolet Near Infrared Optical Northern
Survey (UNIONS) to validate the catalogue used for cosmological analysis
(\citeproc{ref-unions_bmodes}{Daley et al., 2026}). We are also planning
to use this code in the Roman High Latitude Imaging Survey (HLIS). In
addition to its current usage in science collaborations, we provide unit
tests that not only validate the implementation but also validate the
computation mathematically and provide a higher bound for the accuracy
of the code. Fianlly, examples can be found in the code repository that
provide comparison against alternative approaches and implementations.
They show that the computation presented here is significantly faster
than existing alternatives.

\section{AI usage disclosure}\label{ai-usage-disclosure}

Artificial Intelligence (AI) has been use to help with documentation,
docstrings and for some of the unit tests.

\section{Acknowledgements}\label{acknowledgements}

The authors acknowledge the support of a grant from the Simons
Foundation (Simons Investigator in Astrophysics, Award ID 620789).

\section*{References}\label{references}
\addcontentsline{toc}{section}{References}

\protect\phantomsection\label{refs}

\end{document}